\documentclass[preprint,showpacs,preprintnumbers,amsmath,amssymb]{revtex4}

\usepackage{graphicx}
\usepackage{dcolumn}
\usepackage{bm}

\begin{document}

\preprint{FT-HEP001}

\title{The $\omega \rho \pi$ coupling in the VMD model revisited}

\author{D. Garc\'{\i}a Gudi\~no and G. Toledo S\' anchez}
\affiliation{Instituto de F\'{\i}sica,  Universidad Nacional Aut\'onoma de M\'exico, AP 20-364,  M\'exico D.F. 01000, M\'exico}%

\date{\today}

\begin{abstract}
We determine the value of the $\omega -\rho- \pi$ mesons coupling ($g_{\omega\rho\pi} $), in the context of the vector meson dominance model, from radiative  decays,  the $\omega \rightarrow 3\pi$ decay width and the $e^+e^- \rightarrow 3\pi$ cross section.
For the last two observables we consider the effect of either a heavier resonance ($\rho'(1450)$) or a contact term. A weighted average of the results from the  set of observables yields  $g_{\omega\rho\pi} =14.7 \pm 0.1$ GeV$^{-1}$ in absence  of those contributions, and $g_{\omega\rho\pi} =11.9 \pm 0.2$ GeV$^{-1}$ or  $g_{\omega\rho\pi} =11.7 \pm 0.1$ GeV$^{-1}$ when including the $\rho'$ or  contact term respectively. The inclusion of these additional terms makes the estimates from the different observables to lay in a more reduced range.
 Improved measurements of these observables and the $\rho'(1450)$ meson parameters are needed to give a definite answer on the pertinence of the inclusion of this last one in the considered processes.  

\end{abstract}

\maketitle

\section{Introduction}

The strong interaction between the $\omega$, $ \rho$ and $\pi$ mesons  can be encoded in a single parameter, denoted by $g_{\omega \rho \pi}$. These mesons
are produced in experiments devoted to the hadronic production from electron-positron annihilation and tau decays \cite{cmd200,cmd2new,tau,kloe,aulchenko,achasov02,achasov03,pdg}. The increasing experimental accuracy allows to compare the determined strong coupling from different processes to verify their process independent value. Such coupling may have implications in other observables like the muon magnetic moment \cite{benayoung-2a,benayoung-2b}. 
The direct determination of the magnitude of $g_{\omega \rho \pi}$ would require the observation of the $\omega$ decaying into the others, which is not allowed, since there is not enough phase space for the three particles to be on the mass shell. Therefore, it must be extracted by indirect means, for example,  from the above mentioned annihilation process and tau decays. A theoretical framework is required to describe such processes and link the parameters to the physical states,  a Chiral approach based on the low energy symmetries of Quantum Chromodynamics (QCD) and the so called vector meson dominance model (VMD) are able to describe them. Although they have different spirit, they both manage to resume the low energy manifestation of the strong interaction.

In this work we determine the value of $g_{\omega \rho \pi}$, in the context of VMD, from several processes: i) radiative decays of the form $V \rightarrow \pi \gamma $ ($V$: vector meson), ii) the  $\omega \rightarrow 3\pi$ decay and iii) the $e^+e^- \rightarrow 3\pi$ cross section. In the last two cases we explore the possible corrections due the presence of the heavier resonance $\rho '(1450)$ and a $\omega \rightarrow 3\pi$ contact term. At the end we discuss and compare our results with other estimates.

The VMD Lagrangian including the $\rho$, $\pi$ and $\omega$ mesons can be set as:
\begin{eqnarray}
{\cal L}&=& g_{\rho\pi\pi} \epsilon_{abc} \rho_\mu^a \pi^b \partial^\mu \pi^c 
+g_{\omega\rho\pi}\delta_{ab}\epsilon^{\mu\nu\lambda\sigma}\partial_\mu \omega_\nu \partial_\lambda \rho_\sigma^a  \pi^b \nonumber \\
&+&g_{3\pi} \epsilon_{abc}   \epsilon^{\mu\nu\lambda\sigma}\omega_\mu \partial_\nu \pi^a  \partial_\lambda  \pi^b  \partial_\sigma \pi^c +
\frac{e m_V^2}{g_V}V_\mu A^\mu + ...
\label{lagrangian}
\end{eqnarray}
This Lagrangian exhibits only the relevant pieces for this work and should be part of any effective Lagrangian describing these mesons. Terms with higher
derivatives and additional terms which allow to preserve gauge invariance are not shown \cite{klz}. We have made explicit the notation regarding the couplings and the corresponding fields and, in the last term, $V$ refers in general to vector mesons and $A^\mu$ refers to the photon field. Here $g_V=2 \alpha \sqrt{\pi m_V/(3\Gamma(V \rightarrow e^+ e^-))}$, although in general it corresponds to the inclusive leptonic decay  $\Gamma(V \rightarrow l^+ l^-)$ ($l=e,\mu,\tau$). 
The couplings, in this context, are free parameters to be determined from experiment. Relations between them and those coming from low energy theorems can be drawn by building models which incorporate vector mesons into the chiral symmetric lagrangians \cite{kura,kay,fujiwara,kura1,xral,prades94}. In the following we will determine such couplings from the experimental data and  draw some comparisons whenever possible.

\section{Radiative decays}

The $g_{\omega \rho \pi }$ coupling can be obtained from vector mesons radiative decays considering that the photon emission is mediated by a neutral vector meson \cite{vmdrelation}  ( See Fig. \ref{wradfig}). Let us consider the following three decays:

\begin{figure}
\begin{center}
\includegraphics[scale=0.35]{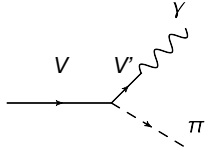}
\end{center}
\caption{Radiative decay of vector mesons. }
\label{wradfig}
\end{figure}

i) The $\omega \rightarrow \pi \gamma $ decay. It is driven by the  $\omega \rightarrow \pi \rho \rightarrow \pi \gamma $ process. Thus, the couplings from both descriptions are related by:
\begin{equation}
g_{\omega \rho \pi }=g_{\omega \pi \gamma}\frac{g_\rho}{e}.
 \end{equation}

ii) The $\rho \rightarrow \pi \gamma $ decay. It is driven by the $\rho \rightarrow \pi \omega (\phi) \rightarrow \pi \gamma $ processes. The presence of the $\phi$ meson  channel can make a sizably contribution \cite{lichard94}, and the relation between the couplings from both descriptions takes the following form:
\begin{equation}
g_{\omega \rho \pi }=\frac{g_\omega}{e} \left(g_{\rho \pi \gamma} - g_{\phi\rho\pi}\frac{e}{g_{\phi}} \right).
\end{equation}
For the above cases, we can compute the $g_{V \pi\gamma}$ coupling using that the radiative decay width is $\Gamma(V \rightarrow \pi\gamma)=g_{V \pi\gamma}^2(M_V^2-m_\pi^2)^3/(96\pi m_V^3)$ and the corresponding  experimental value \cite{pdg}.
 Here, as an approach, we use $|g_{\phi\rho\pi}|=0.86 \pm 0.01$ GeV$^{-1}$ obtained by considering the decay width of the $\phi \rightarrow 3 \pi$ to be fully accounted by the $\rho \pi$ channel (contributions from other channels are relatively smaller \cite{kloe}) \cite{phirhopi1,phirhopi2,phirhopi3}, with the $\phi$ described  similarly to the $\omega$ meson. 
 
 iii) The $\pi^0 \rightarrow \gamma \gamma$ decay.  There are two ways the process can go through: $\pi^0 \rightarrow \rho \omega (\phi)\rightarrow \gamma \gamma$. Taking a global decay constant, $g_{ \pi\gamma\gamma}$, the width can be written as $\Gamma(\pi \rightarrow \gamma\gamma)=g_{ \pi\gamma\gamma}^2 \alpha^2 \pi m_\pi^3/4$.  On the other hand, the destructive interference between the $\rho-\omega$ and the  $\rho-\phi$ channel \cite{lichard11} requires the couplings to be related by :  
\begin{equation}
 |g_{ \pi\gamma\gamma}|=\frac{2}{ g_\rho}\left( \frac{|g_{\omega \rho \pi }|}{ g_\omega} -  \frac{|g_{\phi \rho \pi }|}{ g_\phi}\right).
 \end{equation}

In Table \ref{gwrp}  we show the numerical values for $|g_{\omega \rho \pi }|$ obtained from the above processes. The two most precise determinations are not in agreement with each other, at this stage we can not point out to the source of such deviation.
Neglecting the small correlations induced by $g_V$, we can compute  a weighted average, which yields $11.9 \pm 0.2 $ with the error dominated by the most precise  $\omega \rightarrow \pi^{0} \gamma$ channel.
A standard average gives $g_{\omega \rho \pi }=12.2 \pm 1.3$ GeV$^{-1}$, with errors added in quadratures and is dominated by the uncertainty in the $\rho^{0} \rightarrow \pi^{0} \gamma$ decay width. In the following we will refer to the weighted average from radiative decays as VMDr.
 
\begin{table}
\begin{tabular}{|cc|} 
\hline
     Decay & $|g_{\omega \rho \pi}|$ [GeV$^{-1}$]  \\ \hline 
     $\rho^{-} \rightarrow \pi^{-} \gamma$ & $11.3 \pm 0.9$  \\ 
      $\rho^{0} \rightarrow \pi^{0} \gamma$ & $13.1 \pm 0.9$ \\ 
      $\omega \rightarrow \pi^{0} \gamma$ & $11.4 \pm 0.2$ \\ 
      $\pi^{0} \rightarrow \gamma \gamma$ & $12.8 \pm 0.3$ \\ 
      Weighted Average & $11.9 \pm 0.2$ \\
\hline
\end{tabular}
\caption{Determination  of $|g_{\omega \rho \pi}|$ from radiative decays.}
\label{gwrp}
\end{table}

 As a way of comparison, let us get the expected value from the agreement between VMD and low energy theorems \cite{fujiwara,lowetheorem1,lowetheorem2,lowetheorem3,anomaly1,anomaly2} for the $\pi \rightarrow \gamma \gamma$ decay. Considering only the $\rho-\omega$ channel (the  $\rho-\phi$ channel makes a small effect), the couplings from both descriptions are related by $|g_{\omega\rho\pi}| = |g_{\rho}g_{\omega}/8\pi^2f_\pi|= 11.5$  GeV$^{-1}$, which is in agreement with  the value from radiative decays. If, in addition, we  impose the universality condition ($g_\rho=g_{\rho\pi\pi}$) and SU(3) symmetry  ($3g_\rho=g_\omega$) then $|g_{\omega\rho\pi}| =  |3g_{\rho\pi\pi}^2/8\pi^2f_\pi|=14.4$ GeV$^{-1}$. From this result and using the KSFR\cite{ksfr1,ksfr2} relation ($g_{\rho\pi\pi}=m_\rho/(\sqrt{2} f_\pi$)) it takes the form $|g_{\omega\rho\pi}| =  |3m_{\rho}^2/16\pi^2f_\pi^3|=14.2$ GeV$^{-1}$, where $f_\pi=0.093$ GeV. In these cases there is a significant deviation from the radiative estimate.

\section{The $\omega \rightarrow 3\pi$ decay}

This decay was suggested  to be dominated by the $\omega \rightarrow \rho\pi \rightarrow 3\pi$ process \cite{gell}, and the experimental evidence has provided support to it \cite{cmd200,cmd2new,kloe,aulchenko,achasov02,achasov03}. Additional contributions like the $\rho'$ and contact term are not excluded, which may impact on the determination of the $g_{\omega \rho \pi }$. An analysis of the possible presence of a contact term in this decay was made  in reference \cite{kura}, concluding, at that time, that ¬the existence and magnitude of the contact term can be extracted neither from theory, nor experiment¬. The increasing experimental accuracy on the determination of the $\omega$ decay width \cite{pdg}, will be used here to determine the $g_{\omega \rho \pi }$ coupling in the different scenarios and compare with the corresponding value from other observables.

Let us set our notation for the process $\omega(\eta,q) \rightarrow  \pi^{+}(p_1) \pi^{-}(p_2) \pi^{0}(p_3)$, where the $p_i$ are the corresponding 4-momenta and $\eta$ is the $\omega$ meson polarization. 

\begin{figure}
\begin{center}
\includegraphics[scale=0.35]{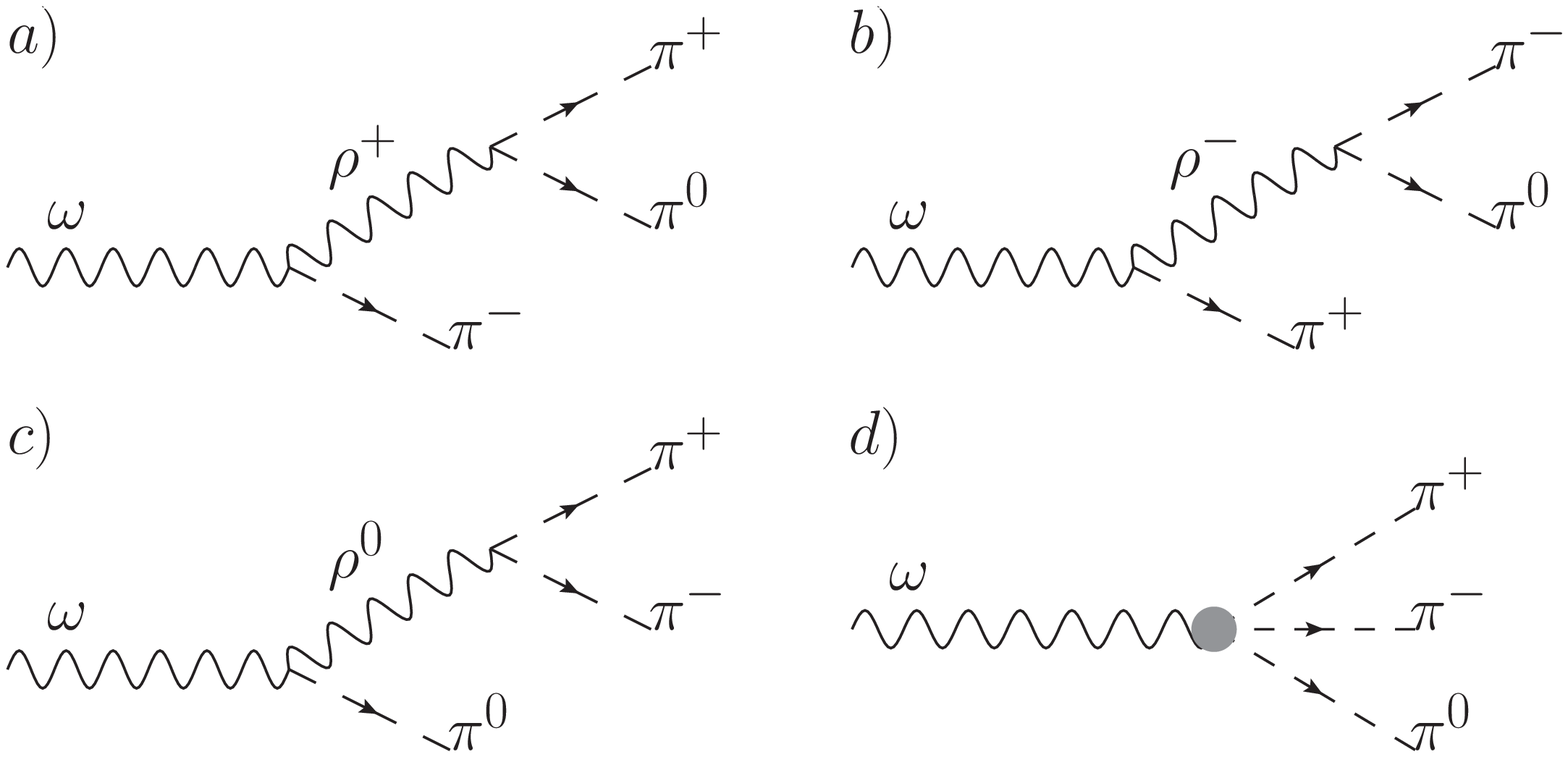}
\end{center}
\caption{$\omega \rightarrow 3\pi$ process. The contribution from the $\rho$ channel  ($a,b,c$) and higher order contributions including the contact term ($d$). }
\label{w3pifig}
\end{figure}

The contributions to the amplitude from the $\rho$ channel (Fig. \ref{w3pifig}a-c) and the contact term (Fig. \ref{w3pifig}d ) can be set as follow:
\begin{equation}
\mathcal{M}_D = \imath \epsilon _{\mu \alpha \beta \gamma} \eta^{\mu} p_1^{\alpha }p_2^{\beta }p_3^{\gamma } \mathcal{A},
\label{ampf3pi}
\end{equation} 
where $\mathcal{A}$ is given by:
\begin{eqnarray}
\mathcal{A} &=& 6g_{3 \pi} +  2g_{\omega \rho \pi } g_{\rho \pi \pi }\left(D^{-1}[\rho^0,p_1+p_2]+\right.
\nonumber \\
&&\left. D^{-1}[\rho^+,p_1+p_3]+D^{-1}[\rho^-,p_2+p_3]\right),
\label{Apdem}
\end{eqnarray}
where, $D[\rho, Q]=Q^2-m_{\rho}^2+i m_{\rho }\Gamma _{\rho }$ and the factors of 6 and 2 in  $\mathcal{A}$ come from the cyclic permutations and momentum conservation used to bring the amplitude into the current form.
 The coupling $g_{\rho \pi \pi }=5.95\pm 0.02$  is fixed by the decay width of $\rho \rightarrow \pi\pi$, $\Gamma_\rho=149.1 \pm 0.8$ MeV.  Using these values and $g_{\omega \rho \pi }$  from radiative decays we can check that the prediction for the width, without taking into account the contact term ($g_{3 \pi} =0$), is $\Gamma^{\rho}_{\omega \rightarrow 3 \pi} = 4.4 \pm 0.2$ MeV, which is  $58 \%$ of the experimental value ($\Gamma^{exp}_{\omega \rightarrow 3 \pi} = 7.56 \pm 0.13$ MeV \cite{pdg}). The correction by using an energy dependent width of the $\rho$ is negligible compared with the error bars and the radiative corrections have been also estimated to be negligible \cite{cmd200}.\\

In order to  reach the 100\% of the experimental width it may be necessary either to increase the coupling constant value from  radiative decays up to $15.7$ GeV$^{-1}$  or to keep such value  and include additional contributions as an effective contact term. 
 
A blind inclusion of the contact term, to account for the observed decay width,  would require : $ g_{3 \pi}= -62 \pm 7 $ or $+409 \pm 10$ GeV$^{-3}$.
However, the proper inclusion is not arbitrary. It is strongly related to the Wess-Zumino-Witten anomaly (WZW) \cite{anomaly1,anomaly2}. In the following we illustrate this point:

The WZW anomaly fixes the amplitude  for the $\gamma^* \rightarrow 3\pi$ decay to be (following the same notation as in Eqn. (\ref{ampf3pi})):
 \begin{equation}
  e\mathcal{A}^{WZW}=\frac{\alpha}{\pi f_\pi^3}.
 \end{equation}

In the VMD approach,  this decay can be produced through the $\omega$ into $\rho \pi$ decay channel, followed by the break down of the $\rho$ into another two pions.
The decay amplitude is similar to Eqn.  (\ref{Apdem}) via the $\rho$ meson  (with $g_{3 \pi}=0$), at $Q^2=0$ and $\Gamma _{\rho}=0$ in the propagators, times a global factor $e/g_\omega$ accounting for the photon-$\omega$ coupling:
\begin{equation}
  \mathcal{A}^\rho=\frac{6e}{g_\omega}\frac{ g_{\omega \rho \pi}g_{\rho \pi \pi}}{m_\rho^2}= \frac{3}{2}\frac{e}{4 \pi^2 f_\pi^3}=\frac{3}{2} \mathcal{A}^{WZW},
\end{equation}
 that is three halves of the total amplitude as obtained from the Chiral anomaly \cite{anomaly1,anomaly2}, while respecting the KSFR relation \cite{rudaz,cohen}.
Therefore, the equivalence between both descriptions requires {\em all the remaining contributions to be collected in the $g_{3\pi}$ coupling}. Then,
\begin{equation}
\frac{-1}{2}\mathcal{A}^{WZW}= \frac{6e g_{3 \pi} }{g_\omega}  \  
\rightarrow g_{3 \pi}=-\frac{g_{\rho\pi\pi}}{16\pi^2f_\pi^3}=- 47 \ GeV^{-3}, \label{lowrelation}
\end{equation}

\noindent where we have made use of the relationship among the couplings as discussed in the previous section. Considering this value in the $\omega \rightarrow 3 \pi$ decay, we  get $\Gamma_{\omega \rightarrow 3 \pi} = 6.8 \pm 0.2$ MeV.
Note that Eqn. (\ref{lowrelation}) is close to  $ g_{3 \pi}= -62 \pm 7 $, obtained blindly to  account for the experimental decay width.\\
This procedure establishes the proper way to include additional contributions in the VMD framework, while keeping the agreement with the low energy theorems. Other approaches also find that there is a need of the contact term \cite{pich03} for a better description of the experimental width.

\subsection{The $\rho'$ channel}
\begin{figure}
\begin{center}
\includegraphics[scale=0.35]{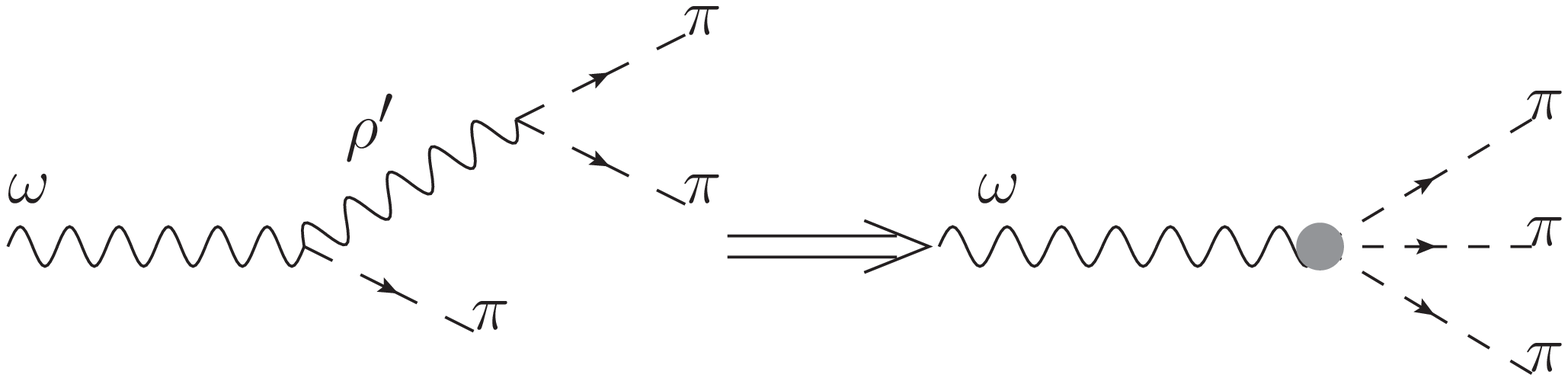}
\end{center}
\caption{$\rho'(1450)$ as an  effective contact term. }
\label{contacto}
\end{figure}

 So far, we have considered the $\rho$ channel and a WZW fixed contact term as the only ways  the decay can go through. However, decays via radial excitations may be also important, provided the mass suppression factor is not extremely large compared to the energies involved in the process.
The  $\rho'(1450)$ meson  ($m_{\rho'}=1465$ MeV and $\Gamma_{\rho'}=400 \pm 60$ MeV), satisfies this condition for the $\omega$ decay regime. Let us explore the role of such contribution:  
 The heavy mass of the  $\rho'$   allows to simplify its propagator  leading to identify the global coupling as an effective contact term (Fig.\ref{contacto}), in full analogy to Eqn. (\ref{lagrangian}):
\begin{equation}
 |g'_{3 \pi} | \approx  \frac{g_{\omega \rho' \pi } g_{\rho' \pi \pi }} {m^2_{\rho'}}.
\label{cv}
\end{equation} 
The couplings involved in the right hand side are not settled, neither in the theoretical side nor experimentally \cite{tau,prime}. 
In order to make an estimate of their magnitudes,  we assume that  $g_{\omega \rho' \pi }/ g_{\rho' \pi \pi }=g_{\omega \rho\pi }/ g_{\rho \pi \pi } = 2$ GeV$^{-1}$. 
This relation is based on the expectation that the radial excitation information of the vector meson cancels out when computing the ratio between processes and considering the central value for  $g_{\omega \rho \pi}$ from radiative decays. Studies on the value of $|g_{\omega \rho' \pi }|$ have found it to lay in the interval 10-18 GeV$^{-1}$ \cite{prime}. Under these assumptions we get  $|g'_{3 \pi}|  \approx 46 \pm 23$ GeV$^{-3}$. 
We have evaluated the deviations from this value in Eqn. (\ref{cv}) due to momentum and width dependence of the propagators, which combined produce an increase of 6\%. Thus, our estimate for the coupling is $|g'_{3 \pi}| =49 \pm 24$ GeV$^{-3}$.\\
Note that the $\rho'$  contribution can not be taken simultaneously with the contact term, Eqn. (\ref{lowrelation}), since this last was determined as if all additional vectors an point contributions were included. Here we limit ourselves to show at what extend the role of the $\rho'$ contribution becomes important.
Recomputing  the total width for $\omega \rightarrow 3\pi$ including the $\rho'$ term yields that, although the  contribution itself turns out to be only $ 4\% \pm 5 \%$ ($0.3 \pm 0.4$ MeV) of the total width, its interference with the $\rho$ channel becomes $29\% \pm 14 \%$ ( $2.2 \pm 1.1 $ MeV), making a global result closer to the experimental value.

In Table \ref{gw3pi}, we have collected a set of values for the contact coupling computed in the literature from different approaches. \cite{rudaz} uses the VMD approach being consistent with low-energy theorems. \cite{cesareo}  Extended the previous idea by including an infinite number of radial excitations.  \cite{kura,kura1} works within a framework of a minimal embedding of vector mesons in a chiral effective langrangian. \cite{kay} uses an extension to the chiral lagrangian adding spin-1 fields, and our results obtained from different approaches.
It can be argued that given the different nature of these approaches a direct comparison between them is meaningless. However, we consider that it is interesting to quote their magnitudes as a way to exhibit that, besides the model dependence, there seems to be a tendency to favor values in the range 30-60 GeV$^{-3}$, the largest difference coming from the estimates at  \cite{kura,kura1}.

\begin{table}
\begin{tabular}{|cc|} 
\hline
     Reference & $|g_{3 \pi}|$ [GeV$^{-3}$]  \\ \hline 
    Rudaz, Cohen \cite{rudaz,cohen} & 47   \\ 
   Dominguez \cite{cesareo} & $29 \pm 3$ \\
    Kuraev et al \cite{kura,kura1} & 123  \\ 
    Kaymakcalan et al \cite{kay} & 37  \\ 
    This work from $\Gamma(\omega \rightarrow 3\pi)$ & $65 \pm 7$  \\  
    This work from $\rho'$& $49 \pm 24$  \\ 
\hline
\end{tabular}
\caption{Determination  of  $|g_{3 \pi}|$-like terms from several approaches. See text for details.}
\label{gw3pi}
\end{table}

\section{The $e^+e^- \rightarrow \omega \rightarrow 3\pi$ cross section}
\begin{figure}
\vspace*{.5cm}
\begin{center}
\includegraphics[scale=0.35]{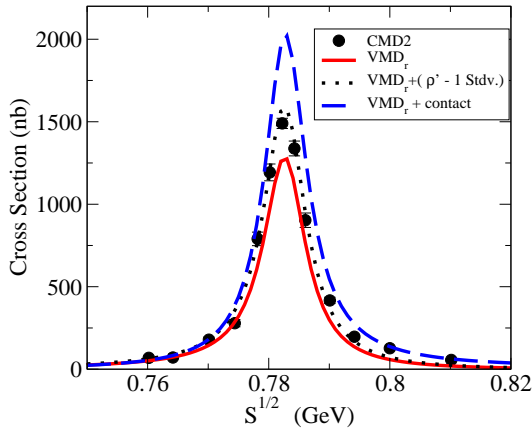}
\end{center}
\caption{ $e^+e^- \rightarrow \omega \rightarrow 3\pi$ cross section. 
Experimental data (circle symbols) and the prediction from VMD using $g_{\omega\rho\pi}$ from radiative decays (solid line), and including either the $\rho'$ term (decreased by 1stdv., dotted line) or the contact term (long dashed line)}
\label{eevsdat}
\end{figure}

Now, we  explore the implications of the contact term or the $\rho'$ in the  $e^+e^- \rightarrow 3\pi$ cross section.
Following the same notation as in the previous section, we can write the amplitude for the $\omega$ channel as follows:
\begin{equation}
\mathcal{M} = \frac{e^2m_{\omega }{}^2}{g_{\omega }}\frac{\bar{v} \gamma^{\mu }u  \epsilon _{\mu \alpha \beta \gamma }p_1{}^{\alpha }p_2{}^{\beta }p_3{}^{\gamma } }
{q^2\left(q^2-m_{\omega }+i m_{\omega  }\Gamma _{\omega }\right)}   \mathcal{A} 
\label{ampee3pi}
\end{equation} 
 where $e$ is the positron electric charge, $m_\omega$ and $\Gamma_\omega$ are the mass and total width of the $\omega$ meson. In Figure \ref{eevsdat} we have plotted the cross section as  function of the center of mass energy. We  show the experimental data from CMD2 \cite{cmd2new} (circle symbols) as a comparison for the VMD prediction when the $\omega$ decay proceeds via the $\rho$ meson  for three cases: \textbf{i}) Using the central value for $g_{\omega\rho\pi}$ coupling determined from the radiative decays (VMD$_r$, solid line),  \textbf{ii}) Including the $\rho'$ contribution (decreased by 1 standard deviation, dotted line) and \textbf{iii}) Including the contact contribution (long dashed line).
 It is clear that the  inclusion of either the $\rho'$ or the contact term increases the central theoretical prediction based only on the $\rho$ channel.  The large uncertainties on the $\rho'$  allows to bring it down to the experimental data  within one standard deviation.  For the fixed contact term, data can be only approached using the VMDr own error bars (not displayed).  
 
\begin{figure}
\vspace*{.5cm}
\begin{center}
\includegraphics[scale=0.35]{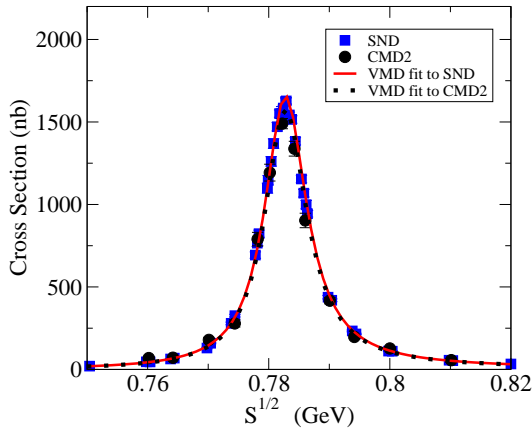}
\end{center}
\caption{ $e^+e^- \rightarrow \omega \rightarrow 3\pi$ cross section. 
Experimental data (circle [CMD2] and square [SND]  symbols) and the corresponding fit by the $\rho$ channel using $g_{\omega\rho\pi}$ as the only free parameter.}
\label{eefit}
\end{figure}

We have performed a fit to the experimental data from CMD2 \cite{cmd2new} and SND \cite{achasov03}. Considering only the $\rho$ channel and leaving $g_{\omega\rho\pi}$ as a free parameter, we get $|g_{\omega\rho\pi}|_{CMD2}=13.1 \pm 0.3$ GeV$^{-1}$ ($\chi^2/npoints\approx 4$) and $|g_{\omega\rho\pi}|_{SND}=13.4 \pm 0.2$ GeV$^{-1}$ ($\chi^2/npoints \approx 1$). The weighted average gives:  $|g_{\omega\rho\pi}|=13.3 \pm 0.2$ GeV$^{-1}$.
In Figure \ref{eefit} we show the experimental data and the corresponding fits. The larger number of data points from SND allows a better fit. The error bars take into account the different values for the mass and width of the $\omega$ meson, taken as the PDG averages or the own experiment determination. 

The same fit procedure is done to determine the coupling upon the inclusion of the contact term, with a fit quality similar to the previous case. We get $|g_{\omega\rho\pi}|_{CMD2}=10.3 \pm 0.1$ GeV$^{-1}$ and $|g_{\omega\rho\pi}|_{SND}=10.6 \pm 0.1$ GeV$^{-1}$. 

If we include the $\rho'$ instead of the contact term, we get $|g_{\omega\rho\pi}|_{CMD2}=10.1 \pm 1.5$ GeV$^{-1}$ and $|g_{\omega\rho\pi}|_{SND}=10.5 \pm 1.6$ GeV$^{-1}$. The large error bars are dominated by the uncertainties of the $\rho'$ parameters.

The weighted averages from both experiments for all the cases, along with the determinations from the other observables, are presented in Table \ref{resfinal}.

\section{Discussion}
We have performed an analysis to obtain the  $g_{\omega\rho\pi} $ coupling in the VMD approach, considering radiative decays,  the $\omega \rightarrow 3\pi$ decay width and the $e^+e^- \rightarrow \omega \rightarrow 3\pi$ cross section. Our global result from all these observables are presented within two possible scenarios  (See Table \ref{resfinal}): 
 
 i) Considering that only the $\rho$ meson channel is present: This corresponds to  the value obtained from radiative decays, the value which reproduces the experimental decay width for $\omega \rightarrow 3\pi$ and the value obtained from the fit to the experimental $e^+e^- \rightarrow \omega \rightarrow 3\pi$ cross section \cite{cmd2new,achasov03}. The weighted average (Table \ref{resfinal}, column 2) is  $|g_{\omega\rho\pi}|=14.7 \pm 0.1$ GeV$^{-1}$.
  
 ii) Considering either the $\rho'$ or the contact term, in addition to the $\rho$: In this case we average the value obtained from radiative processes, the values from the decay width for $\omega \rightarrow 3\pi$ and the fits to $e^+e^- \rightarrow \omega \rightarrow 3\pi$ cross section, obtained upon the inclusion of the contact term fixed by the equality condition between VMD and low energy theorems, or upon the inclusion of that produced by the $\rho'$ meson.
 The weighted average (Table \ref{resfinal}, columns 3 and 4) are  $|g_{\omega\rho\pi}|=11.7 \pm 0.1$ GeV$^{-1}$ and $|g_{\omega\rho\pi}|=11.9 \pm 0.2$ GeV$^{-1}$, respectively.
 
The $\rho'$ contribution has uncertainties associated to its corresponding mass, decay width and couplings. This last  was based on an {\em ansatz} to relate the couplings of the $\rho$ with its radial excitation. Further experimental information is required to determine the validity of such approach and thus the pertinence of including this contribution. The inclusion of the contact term in the $\omega \rightarrow 3\pi$ decay is consistent with the low energy theorems while fulfilling the KSFR relation.
 
The results in the first scenario are spread out in the range 11.9 GeV$^{-1}$ to 15.7 GeV$^{-1}$ (central values). The addition of other contributions to the amplitude reduces the discrepancy between the coupling values as determined from different observables. In particular they favor values around 11.9 GeV$^{-1}$.

We can compare our result with estimates based on different approaches: 
Ref.~\cite{lutz09}  obtains 7.35 GeV$^{-1}$ using a counting scheme for flavor-SU(3) systems of Goldstone bosons and light vector mesons.
 Some estimates based on QCD sum rules, Refs.~\cite{Khathimovsky85} and \cite{Margvelashvili88} obtain 9 GeV$^{-1}$,  while Refs.~\cite{Margvelashvili88,sumrules1,sumrules2,sumrules3,lublinsky97} are in the range 15 GeV$^{-1}$ to 17 GeV$^{-1}$. A similar value  of 16 GeV$^{-1}$ is found in Ref.~\cite{cesareo} by the inclusion of an infinite number of radial excitations, and in Refs.~\cite{su3a} and \cite{su3b} under the SU(3) symmetry, the symmetry breaking effect have been also considered in Refs.~\cite{su3break1} and \cite{su3break2}.
Ref.~\cite{khun06} obtains 15.8 GeV$^{-1}$  when only the $\rho$ channel is considered in the $e^+e^- \rightarrow \omega \rightarrow 3\pi$ cross section.
  A quark level linear sigma model, Ref.~\cite{lucio}, gets 10.3 GeV$^{-1}$ to 14.7 GeV$^{-1}$. 
 In the Dyson-Schwinger framework, using the rainbow-lader approximation, Ref.~\cite{maris03}, it is found to be 10.3 GeV$^{-1}$ .

These works favor  two kind of values,  one around 10 GeV$^{-1}$ and another around 16 GeV$^{-1}$. Given the different approaches followed in these works, we can not point out the origin of the differences. As a comparison, in our calculation we have shown that the inclusion of the contact term or a heavier resonance, can bring the estimate of the coupling  from a value as large as $\approx$16 GeV$^{-1}$ to a lower value of $\approx$ 11 GeV$^{-1}$, signing the importance of these contributions. 

The final results  here presented come from  weighted averages and  therefore are dominated by the more precise measurements. The discrepancy between the different results may be an indication of either data inconsistency and/or bad model behavior.
Improved measurements of the observables under consideration and the $\rho'(1450)$ meson parameters are needed to settle the issues above mentioned.

Our approach is based in a generic form of the VMD lagrangian which should be part of any effective lagrangian including the mesons involved. In our analysis we have considered several kinds of observables. Therefore, on those grounds, our results are solid and may be useful to compute other processes.
  
\begin{table}
\begin{tabular}{|c|c|c|c|}
\hline
      & $\rho$ channel& $\rho$ + contact &$\rho$ + $\rho'$ \\ \hline 
     VMD$_r$ & $11.9 \pm 0.2$ &$11.9\pm 0.2$ &$11.9 \pm 0.2$ \\ 
     $\Gamma (\omega \rightarrow 3 \pi)$ & $15.7 \pm 0.1$ &$12.8\pm 0.1$ &$12.6 \pm 1.3$ \\ 
     $\sigma (e^+e^- \rightarrow 3 \pi)$  & $13.3 \pm 0.2$ & $10.5 \pm 0.1$&$10.3 \pm 1.6$ \\ 
\hline
 Average & $ 14.7\pm0.1 $ &$11.7\pm 0.1$ & $11.9 \pm 0.2 $ \\
\hline
\end{tabular}
\caption{The $|g_{\omega \rho \pi}|$ coupling (GeV$^{-1}$) from different scenarios and observables.}
\label{resfinal}
\end{table}

\begin{acknowledgments}
We acknowledge the support of CONACyT,  M\'exico. We thank Dr. G. L\'opez Castro, Dr. Jens Erler and Dr. Peter Lichard for very useful observations.
\end{acknowledgments}



\begin{thebibliography}{50}    

\bibitem{cmd200} R.R. Akhmetshin et al. {\it Phys. Lett. B} {\bf 476}, 33(2000). 

\bibitem{cmd2new} R.R. Akhmetshin, et al. {\it Phys. Lett. B} {\bf 578}, 285(2004).
\bibitem{tau} K. W. Edwards et al.  {\it Phys. Rev. D} {\bf 61}, 072003 (2000).

\bibitem{kloe} F. Ambrosino, et al.  {\it Phys. Lett. B} {\bf 669}, 223(2008).

\bibitem{aulchenko} V. M. Aulchenko, et al {\it Jour. Exp. Theo. Phys.} {\bf 90}, 927(2000)

\bibitem{achasov02}  M. N. Achasov et al.  {\it Phys. Rev. D}  {\bf 65} 032002(2002).
\bibitem{achasov03}  M. N. Achasov et al.  {\it Phys. Rev. D} {\bf 68} 052006 (2003).

\bibitem{pdg}K. Nakamura et al. (Particle Data Group), {\it J. Phys. G} {\bf 37}, 075021 (2010)

\bibitem{benayoung-2a}M. Benayoun, P. David, L. DelBuono, O. Leitner, {\it Eur. Phys. Jour C} {\bf 65}, 211(2010).

\bibitem{benayoung-2b}M. Benayoun, P. David, L. DelBuono, O. Leitner, {\it Eur. Phys. Jour C}  {\bf 68}, 355(2010).


\bibitem{klz}N. M. Kroll, T.D. Lee and B. Zumino, {\it Phys. Rev.} {\bf 157} 1376(1967). 


\bibitem{kura} E. A. Kuraev, Z. Silagadze, {\it Phys. Atom. Nucl.} {\bf 58}, 1589(1995). 
\bibitem{kay}O. Kaymakcalan, S. Rajeev, J. Schechter, {\it Phys. Rev. D} {\bf 30}, 594(1984).
\bibitem{fujiwara} T. Fujiwara et al. , {\it Prog. Theor. Phys.} {\bf 73}, 92(1985).
\bibitem{kura1} E.A. Kuraev, Z.K. Silagadze , {\it Phys. Lett. B} {\bf 292}, 377(1992). 
\bibitem{xral} J. Gasser, G. Ecker,  A. Pich and E. de Rafael,  {\it Nucl. Phys. B} {\bf 321}, 311(1988).
\bibitem{prades94}Joaquim Prades,Ê{\it Z. Phys. C} {\bf 63}, 491 (1994)

\bibitem{vmdrelation}M. Gell-Mann and F. Zachariasen,  {\it Phys. Rev.} {\bf 124}, 953(1961).

\bibitem{lichard94}P. Lichard,  {\it Phys. Rev. D} {\bf 49}, 5812(1994).

\bibitem{phirhopi1} For other determinations see M. N. Achasov et al. {\it Phys. Rev. D} {\bf 63}, 072002(2001). 
\bibitem{phirhopi2}M. Benayoun and H. B. O'Connell,  {\it Eur. Phys. Jour C} {\bf 22}, 503(2001).
\bibitem{phirhopi3}A. Flores Tlalpa and G. Lopez Castro {\it Phys. Rev. D} {\bf 77} 113011(2008). 

\bibitem{lichard11}See P. Lichard, {\it  Phys. Rev. D} {\bf 83} 037503 (2011) for the role of other channels.

\bibitem{lowetheorem1}S.Adler, B.W.Lee, S.Treiman, A.Zee, {\it Phys. Rev. D} {\bf 4}, 3497(1971).
\bibitem{lowetheorem2}M.V. TerentÕev, {\it JETP Lett.} {\bf14}, 40(1971).
\bibitem{lowetheorem3}M.V. TerentÕev, {\it Phys. Lett. B} {\bf38}, 419(1972).

\bibitem{ksfr1} K. Kawarayabashi  and M. Suzuki , {\it Phys. Rev. Lett.} {\bf16}, 255(1966). 
\bibitem{ksfr2}Riazuddin and Fayyazuddin, {\it Phys. Rev.} {\bf 147}, 1071(1966).

\bibitem{anomaly1}J. Wess , B. Zumino , {\it Phys. Lett. B} {\bf 37}, 95(1971).
\bibitem{anomaly2}E. Witten , {\it Nucl. Phys. B} {\bf 223}, 422(1983). 

\bibitem{gell} M. Gell-Mann, D. Sharp, W. G. Wagner, {\it Phys. Rev. Lett.} {\bf 8}, 261(1962) .


\bibitem{rudaz} S. Rudaz, {\it Phys. Lett. B} {\bf 45}, 281(1984). See references therein for earlier works.
\bibitem{cohen}T. D. Cohen {\it Phys. Lett. B} {\bf 233}, 467(1989)
\bibitem{pich03}ÊSee for example, Pedro D. Ruiz-Femen\'{\i}a, A. Pich and J. Portol\'es,Ê{\it JHEP} {\bf 07}, 003(2003).

\bibitem{prime}  N. N. Achasov and A. A. Kozhevnikov, {\it Phys. Rev. D} {\bf 62}, 117503 (2000) and references therein.

\bibitem{cesareo} C. A. Dominguez, {\it Mod. Phys. Lett. A} {\bf 12}, 983(1987).

\bibitem{khun06}ÊH. Czyz, A. Grzelinska, J.H. Kuhn, and G. Rodrigo, {\it Eur. Phys. Jour. C} {\bf 47}, 617(2006).

\bibitem{lutz09}ÊS. Leupold and M.F.M. Lutz,Ê{\it Eur. Phys. Jour. A}{\bf 39}, 205(2009).
\bibitem{Khathimovsky85}V. M. Khathimovsky, {\it Yad. Fiz.} {\bf 41}, 519 (1985).
\bibitem{Margvelashvili88}M. V. Margvelashvili and M. E. Shaposhnikov, {\it Z. Phys. C} {\bf 38}, 467 (1988).

\bibitem{sumrules1}V.L. Eletsky, B.L. Ioffe, Ya.I. Kogan, {\it Phys. Lett. B} {\bf 122} 423(1983).
\bibitem{sumrules2}S. Narison and N. Paver, {\it Z. Phys. C} {\bf 22}, 69(1984).
\bibitem{sumrules3}V. M. Braun and I. E. Filyanov, {\it Z. Phys. C} {\bf 44}, 157 (1989).

\bibitem{lublinsky97}M. Lublinsky, {\it Phys. Rev. D} {\bf 55}, 249(1997).

\bibitem{su3a}P. Rotelli and M.D. Scadron, {\it Il Nuovo Cimento A} {\bf 15},643 (1973). 
\bibitem{su3b}P.G.O. Freund and S. Nandi, {\it Phys. Rev. Lett.} {\bf 32}, 181(1974).

\bibitem{su3break1} A. Bramon, A. Grau, G. Pancheri {\it Phys. Lett. B} {\bf 344}, 240(1995).
\bibitem{su3break2}M. Hashimoto, {\it Phys. Lett. B} {\bf 381} 465(1996).

\bibitem{lucio} J. L. Lucio et al. {\it Phys. Rev. D} {\bf 61}, 034013(2000).

\bibitem{maris03}Stephen R. Cotanch and Pieter Maris {\it Phys. Rev. D} {\bf 68}, 036006 (2003).



\end{thebibliography}
\end{document}